# Differential Operator in Seizure Detection

Kaushik Majumdar

Systems Science and Informatics Unit, Indian Statistical Institute, 8th Mile, Mysore Road, Bangalore 560059, India; E-mail: kmajumdar@isibang.ac.in

*Abstract* – **Differential operators can detect significant changes in signals. This has been utilized to enhance the contrast of the seizure signatures in depth EEG or ECoG. We have actually taken normalized exponential of absolute value of single or double derivative of epileptic ECoG. Variance operation has been performed to automatically detect seizures. A novel method for determining the duration of seizure has also been proposed. Since all operations take only linear time, the whole method is extremely fast. Seven novel parameters have been introduced whose patient specific thresholding brings down the rate of false detection to a bare minimum. Results of implementation on the ECoG data of four epileptic patients have been reported with an ROC curve analysis. High value of the area under the ROC curve indicates excellent detection performance.**

## I INTRODUCTION

Second derivative based Laplacian operator is widely used for edge detection in an image [1]. An *edge* can be characterized by an abrupt change in intensity indicating the boundary between two regions of an image [2]. We have applied the same logic in this letter to detect the boundary between seizure and nonseizure in ECoG signals of epileptic patients which indicates a seizure onset or offset (for automatic seizure detection see [3] – [5]). Earlier first and second derivative of neonatal sleep EEG were used for feature extraction in order to automatically detect the sleep stages [6]. First and second derivative of EEG were also used to extract time domain features for automatic seizure detection in [7].

In the next section we will describe the method. In section 3 data acquisition will be described. In section 4 will contain the results of implementation on depth EEG or ECoG of four epileptic patients. We will use EEG and ECoG interchangeably throughout this letter. The last section contains some concluding remarks.

## II METHOD

In this letter we will be dealing with digital signals only. Derivative is discrete derivative or difference operation. Let $a$, $b$ and $c$ are successive time points. If a spike occurs at $b$ then statistically $x(b) - x(a)$ and $x(c) - x(a)$ both have high numerical value, where $x()$ is the signal. The second derivative $x(a) + x(c) - 2x(b)$ has an even higher numerical value. Whereas these values for the back ground signal won't be much higher. Let us take the transformation $\exp\left(\frac{1}{w}|D^2 x|\right)$, where $D^2$ is the second derivative, $|.|$ is absolute value and $w$ is a positive valued normalization constant. $\exp\left(\frac{1}{w}|D^2 x|\right)$ acts as a spike enhancement filter with respect to the back ground EEG (spike enhancement through appropriate filter for the detection purpose has also been accomplished in [8]). $\exp\left(\frac{1}{w}|Dx|\right)$ too acts the same way, where $D$ denotes the first order derivative. Depending on the data set one gives better results than the other.

Success of variance in seizure detection is well established [9]. In section 4 we will see that the above filtering can significantly improve the seizure detection accuracy by variance. In order to minimize false detection patient specific threshold needs to be set for the following parameters introduced in this paper. For the detail of implementation see the MATLAB programs with elaborate documentation along with supplementary materials in the author's website [10]. The implementation results for particular patients at any given time slot and channel are summarized in Fig. 1 (for single derivative filter or SDF) and Fig. 2 (for double derivative filter or DDF).

(a) Maximum windowed variance (B) of the filtered data (in the above sense).

(b) Maximum windowed variance of absolute value of the data (C).

(c) $|\max(B) - mm(1)|$, where $mm(1)$ is the variance of filtered data in the window next to the window of the maximum variance.

(d) $mm(i)$ is the variance of filtered data at the $i$th window after the maximum window. Number of windows after the maximum is stipulated (typically at 16). $nl(1)$ is the position of the first of the windows with minimum variance among the stipulated number of windows. If $M$ is the position of $\max(B)$ then $N1 = M + window * n(1)$. $M$ has been treated as the onset of seizure in this paper (in majority of the cases it is several seconds after the actual onset) and $N1$ is the offset (usually several seconds after the actual offset). $E$ is an array consisting of windowed variance of the filtered data starting from two windows before $M$ up to $N1$. $x$ is an array consisting of maximum values of $E$. $F$ is another array consisting of values of $E$ which are greater than or equal to $\frac{3}{4}\max(E)$.

And now we are in a position to say that $\lfloor mean(F) - x(1) \rfloor$ is a quantity whose threshold distinguishes between seizure and nonseizure EEG.

(e) $\lfloor std(F) \rfloor$, where $std$ stands for standard deviation.

(f) $DE = \exp\left(\frac{1}{v}\left|D^2 E\right|\right)$, where $v$ is a positive valued normalization constant. $K$ is an array consisting of values of $DE$, which are greater than or equal to $0.999 * \max(DE)$. Length of $K$, whose threshold distinguishes between seizure and nonseizure EEG.

(g) For seizure EEG max($DE$) must lie within an interval.

Let us mention once again that all the thresholding in the above parameters and the interval in (g) are patient specific. In this work duration of seizure has been calculated as described in (d). Each of the operations executes in linear time. The whole method is extremely fast – takes less than 4 second for a one hour long signal with 15.625 second window length and 15.234 second overlap on an Intel Core 2 Duo Processor T8100 (2.1$GHz$/800$MHz$ FSB, 3M L2 cache), Ubuntu machine with 4GB RAM. The implementation was in MATLAB.

## III Data Acquisition

Four medically intractable focal epileptic patients' ECoG data that have been analyzed in this work have been provided by the Seizure Prediction Project of the Albert-Ludwig-Universitat Freiburg, Germany [11]. In order to obtain a high signal to noise ratio (SNR), fewer artifacts and to record directly from focal areas intracranial grid, strip and depth electrodes were utilized. The ECoG data were acquired using Neurofile NT digital video EEG system (It-med, Usingen, Germany) with 128 channels, 256 Hz sampling rate, and a 16 bit analog to digital converter. In all cases the ECoG from only six sites have been analyzed. Three of them from the focal areas and the other three from out side the focal areas. See Table 1 for the patient details. A superset of the patient population has been studied in [12].

## IV Results

*A) Preprocessing*

Since the data were collected over couple of years, the conditions under which the data had been collected are likely to be different from patient to patient. We have performed different preprocessing for different patients for the optimum results. We have chosen the method by trial and error. Gaussian low pass filter, with cut off frequencies either 50 or 100 Hz depending on the patient, has been used to remove muscle contraction artifacts. Montage change from common reference to bipolar has helped to suppress chewing artifacts in patient 4 to some extent. See Table 2 for the details. For patient 4 the three in-focus electrodes have been put in bipolar reference among themselves and three out-focus electrodes have been put in bipolar reference among themselves (Although intensity of seizure decreases due to subtracting one channel from another which may result in detection failure, in this particular case it helped to eliminate artifacts to a large extent while still preserving the strength of the signal, which has turned out to be sufficient for the detection purpose).

*B) Automatic Detection*

The detection algorithm was run on one hour long segments of ECoG of the patients containing one seizure per segment. Window length of 4000 time points (15.625 second) with 3900 (15.234 second) time points overlap (i.e., sliding by 100 points) has been used. Seizure portions were identified by certified epileptologists at the place of origin of the data at each given time slot, but not for individual channels. The algorithm was implemented on each channel to automatically detect onset and offset of a seizure. Onset of the seizure has been taken to be the earliest point detected as onset among all the channels. For offset also the earliest point detected among the channels has been taken to be the offset.

The SDF has been applied on patients 1 and 4. The DDF has been applied on patients 2 and 3. Data of 1 and 4 contain more artifacts or noise than those of 2 and 3. DDF could not detect some of the seizures of patient 1. In relatively artifact free signals DDF usually gave more accurate measure of the seizure onset than SDF. In the ECoG of patient 1 windowed variance could not detect seizures in the preprocessed signals. But after filtering with the SDF all the seizures could be detected by windowed variance.

For the first three patients seizures were detected with cent percent accuracy. Data of the last four out of five seizures of patient 4 were heavily contaminated by chewing artifact. First seizure has been detected nicely on all channels. Second and fourth seizures have been detected on all focal channels. The third and the fifth seizures have been detected in two out of three focal channels and in one the seizure could not be detected although it showed up clearly in the plot (see supplementary materials in [6]). The fifth seizure ECoG of patient 4 required a very special and unique preprocessing (suppression of all values $\geq 0.15$ of the maximum value irregardless of sign), and therefore we will treat it as a failure to detection. For patients 1, 3 and 4 seizure free data for 24 hours were available on which the detection algorithm was run to test for false positives. There were 4 false detections for patient 1, 5 for patient 3 and 0 for patient 4. Since the available seizure ECoG was rather scarce for each patient, thresholds were set at the time of detection. Then the effectiveness of the method was tested by the number of false positives on the 24 hour long seizure free signals. Performance measure has been given in terms of the area under the ROC curve in the next subsection. The average seizure detection time lag is $20.45$ second after the epileptologist determined onset, which is $9.3$ second in [4]. The average seizure offset detection time lag is $26.49$ second after the epileptologist determined offset.

*C) ROC Curve Analysis*

ECoG recording of $28$ hours for patient 1 is available. For patient 2 the recording is of $3$ hours duration (the $24$ hours seizure free data is not available). For patients 3 and 4 it is for $29$ hours each. The receiver operator characteristic (ROC) curve has been plotted as true positive rate (TPR, plotted along the Y-axis) vs. false positive rate (FPR, plotted along the X-axis) following [13] (Fig. 3). For the purpose of the ROC curve plotting we have assigned 5 false detections to patient 2 in 24 hours, which is the highest in the poll of patients under study. For a discussion on missing values conventions see ref. [14]. $FPR = 1 - \text{specificity} = 1 - \frac{TN}{TN+FP}$ and $TPR = \text{sensitivity} = \frac{TP}{TP+FN}$ [13]. The area under the curve is $\approx 0.98$, which indicates excellent identification accuracy.

## CONCLUSION

In this letter the power of differential operators in detecting significant changes in one dimensional signals such as single channel ECoG has been studied. Excellent accuracy has been observed in detecting seizures with a small number of false detections. The method made up of differential operators, exponentiation and variance has turned out to be extremely fast. Seven parameters involving them have been identified whose patient specific threshold can distinguish between seizure and nonseizure signals for a given patient with impressively high detection and low false detection rate. Further improvements in elimination of false detection are possible with multidimensional statistical analysis of these parameters. Wavelet feature detection of the filtered signal followed by appropriate clustering techniques may result in further improvement in reducing the gap between the detection time and the actual onset time.

The current method has shown promising success on ECoG, which is relatively noise free, but not artifact free. The proposed filter can greatly enhance isolated spikes with respect to the back ground and therefore may be a potential tool for spike detection in single cell recordings. It can also suppress low frequency artifacts like those generated by eye blinks. The same is true for low intensity noise. It is yet to be tested for strong event related potential (ERP) detection in scalp EEG. However it is likely to give good results for automatic quake detection in seismological signals. Particularly many low intensity quakes are difficult to detect, yet they contain important information about the inside of our planet. At times they may even be precursor to an impending major earth quake.

ACKNOWLEDGMENT

The author is grateful to the Freiburg Seizure Prediction Project, Freiburg, Germany, for generously providing the ECoG data. He also likes to thank Hinnerk Feldwisch for helping with the data transfer and answering some questions to better understand it. Dipti P. Mukherjee is acknowledged for some helpful suggestions.

**References**

1. D. Marr and E. Hildrath, "Theory of edge detection," *Proc. Royal Soc. London, Series B, Biol. Sc.*, vol. 207, no. 1167, pp. 187 – 217, Feb 1980.

2. X. Wang, "Laplacian operator-based edge detectors," *IEEE Trans. Patt. Anal. Mach. Intel.*, vol. 29(5), pp. 886 – 890, May 2007.

3. A. A. Dingle, R. D. Jones, G. J. Carroll, and W. R. Fright, "A multistage system to detect epileptiform activity in EEG," *IEEE Trans. Biomed. Eng.*, vol. 40(12), pp. 1260 – 1268, Dec 1993.

4. H. Qu, and J. Gotman, "A patient-specific algorithm for the detection of seizure onset in long-term EEG monitoring: Possible use as a warning device," *IEEE Trans. Biomed. Eng.*, vol. 44(2), pp. 115 – 122, Feb 1997.

5. H. C. Lee, W. van Drongelen, A. B. McGee, D. M. Frim, and M. H. Kohrman, "Comparison of seizure detection algorithms in continuously monitoring pediatric patients," *J. Clin. Neurophysiol.*, vol. 24(2), pp. 137 – 146, Apr 2007.

6. V. Krajca, S. Petranek, J. Mohylova, K. Paul, V. Gerla, and L. Lhotska, "Neonatal EEG sleep stages modeling by temporal profiles," In: *LNCS 4739*, R. Moreno-Diaz (ed), Springer-Verlag, Berlin, Heidelberg, 2007, pp. 195 – 201.

7. N. Paivinen, S. Lammi, A. Pitkanen, J. Nissinen, M. Penttonen, and T. Gronfors, "Epileptic seizure detection: A nonlinear viewpoint," *Comp. Meth. Progr. Biomed.*, vol. 79(2), pp. 151 – 159, 2005.

8. D.-M. Ward, R. D. Jones, P. J. Bones, and G. J. Carroll, "Enhancement of deep epileptiform activity in the EEG via 3-D adaptive spatial filtering," *IEEE Trans. Biomed. Eng.*, vol. 46(6), pp. 707 – 716, Jun 1999.

9. P. E. McSharry, T. He, L. A. Smith, and L. Tarassenko, "Linear and non-linear methods for automatic seizure detection in scalp electro-encephalogram recordings," *Med. Biol. Eng. Comp.*, vol. 40, pp. 447 – 461, 2002.

10. K. Majumdar, MATLAB code and supplementary materials, available at http://www.isibang.ac.in/~kaushik, 2009.

11. Freiburg Seizure Prediction Project, Freiburg, Germany, https://epilepsy.uni-freiburg.de/, 2008.

12. R. Aschenbrenner-Scheibe, T. MAiwald, M. Winterhalder, H. U. Voss, J. Timmer and A. Schulze-Bonhage, "How well can epileptic seizures be predicted? An evaluation of a nonlinear method," *Brain*, vol. 126, pp. 2616 – 2626, 2003.

13. W. Chaovalitwongse, L. D. Iasemidis, P. M. Pardalos, P. R. Carney, D.-S. Shiau, J. C. Sackellares, "Performance of a seizure warning algorithm based on the dynamics of intracranial EEG," *Epilepsy Res.*, vol. 64, pp. 93 – 113, 2005.

14. W. J. Krzanowski, and D. J. Hand, *ROC Curves for Continuous Data*, CRC Press, Boca Raton, 2009.

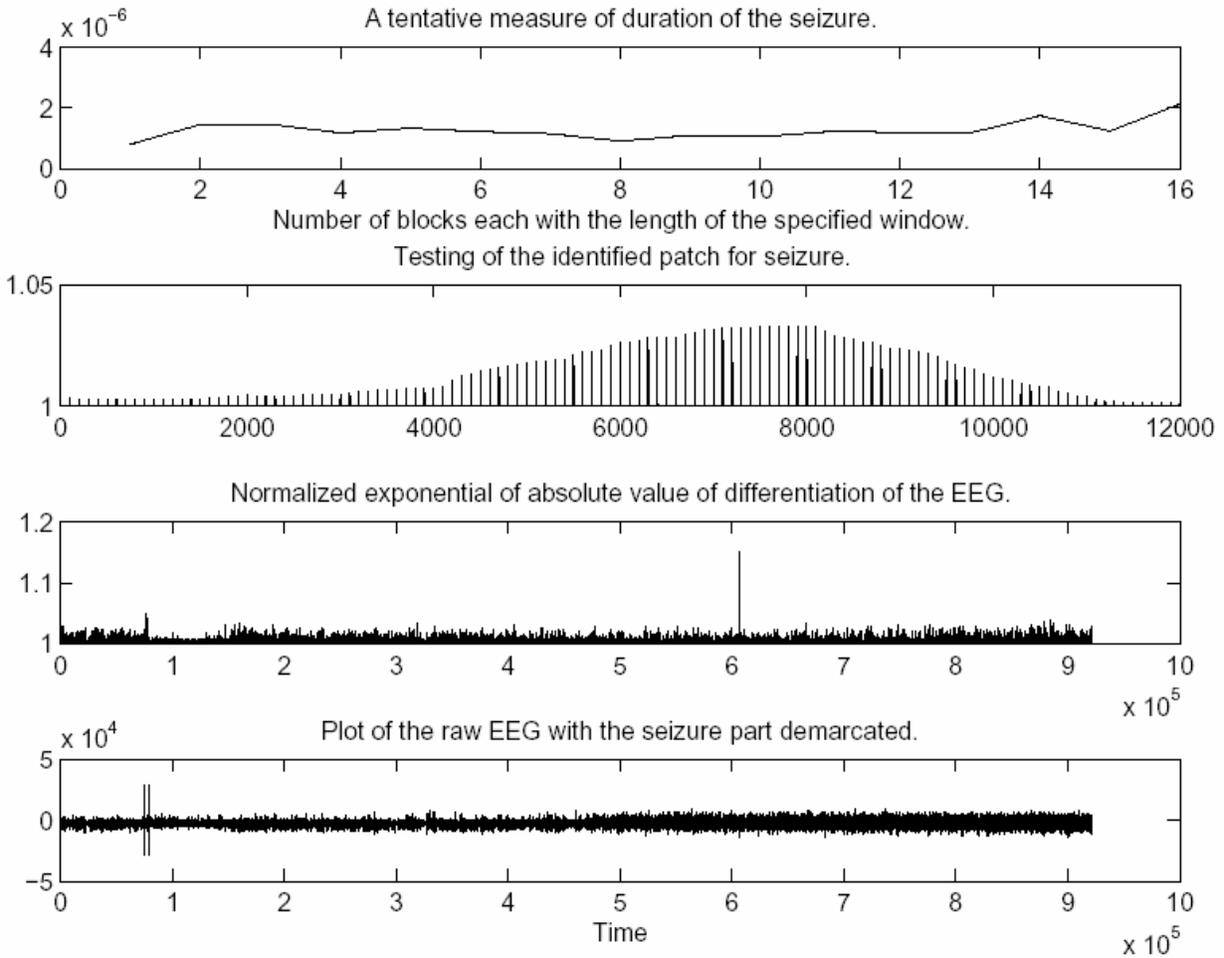

Fig. 1. Automatic detection of seizure and its duration at a single channel for patient 1. Seizure part has been demarcated by parallel vertical lines in the plot of raw EEG at the bottom panel. In the second panel from the bottom the filtered (based on single derivative) signal has been plotted, in which the seizure part is appearing as a distinct pillar like structure with respect to the back ground. The third panel from below plots DE, whose distinct shape corresponds to seizure. The top panel plots mm, which determines a tentative duration of the seizure. Automatic detection is from 74501 to 78501 time points, whereas the actual seizure occurred from 73382 to 78125 time points. 256 time points = 1 second.

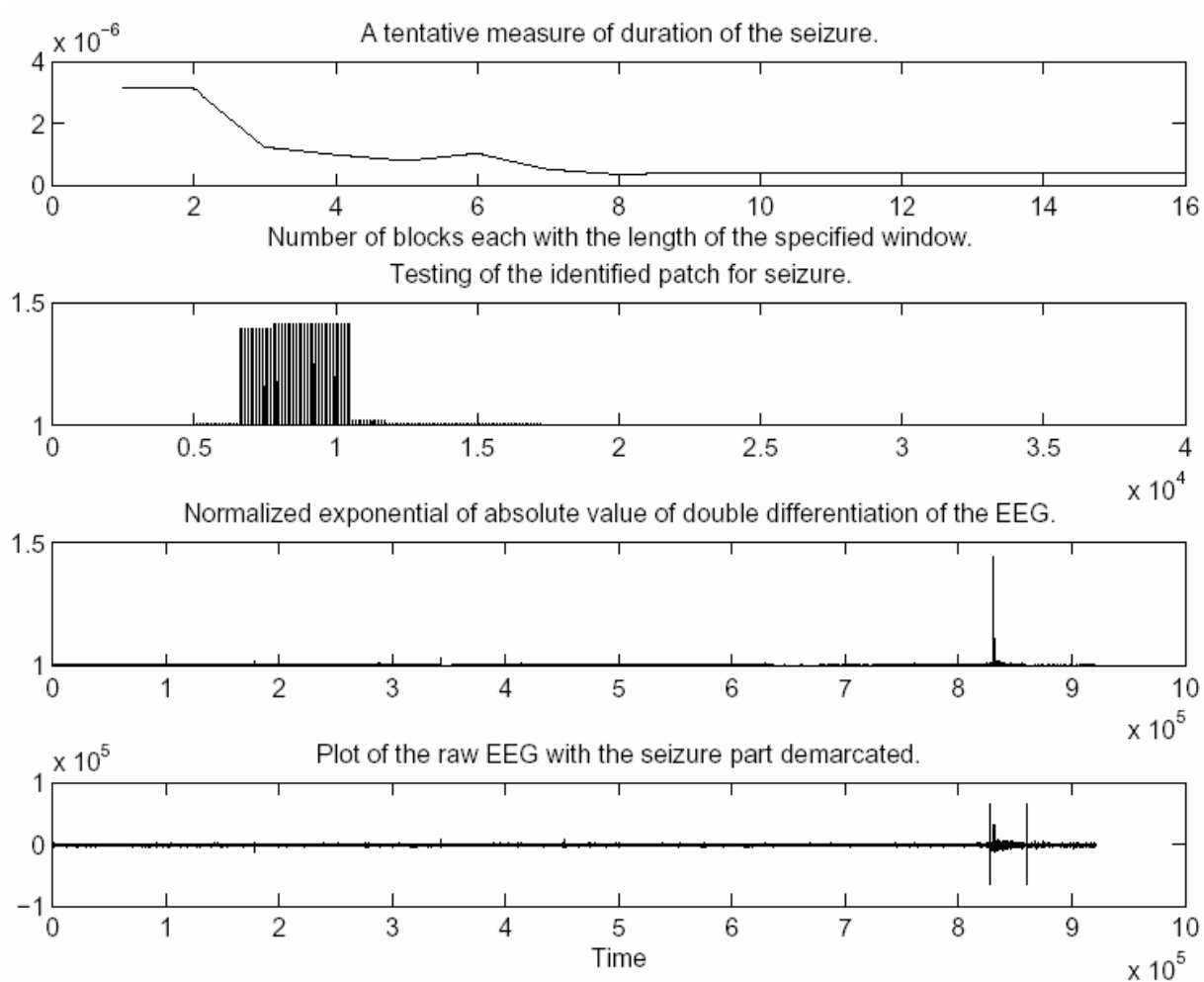

Fig. 2. Same as in Fig. 1, but for patient 2. In this case the filter is based on double derivative of the signal. Automatic detection is from 827901 to 859901 time points, whereas the seizure has been identified by epileptologist is from 819373 to 857094 time points. 256 time points = 1 second.

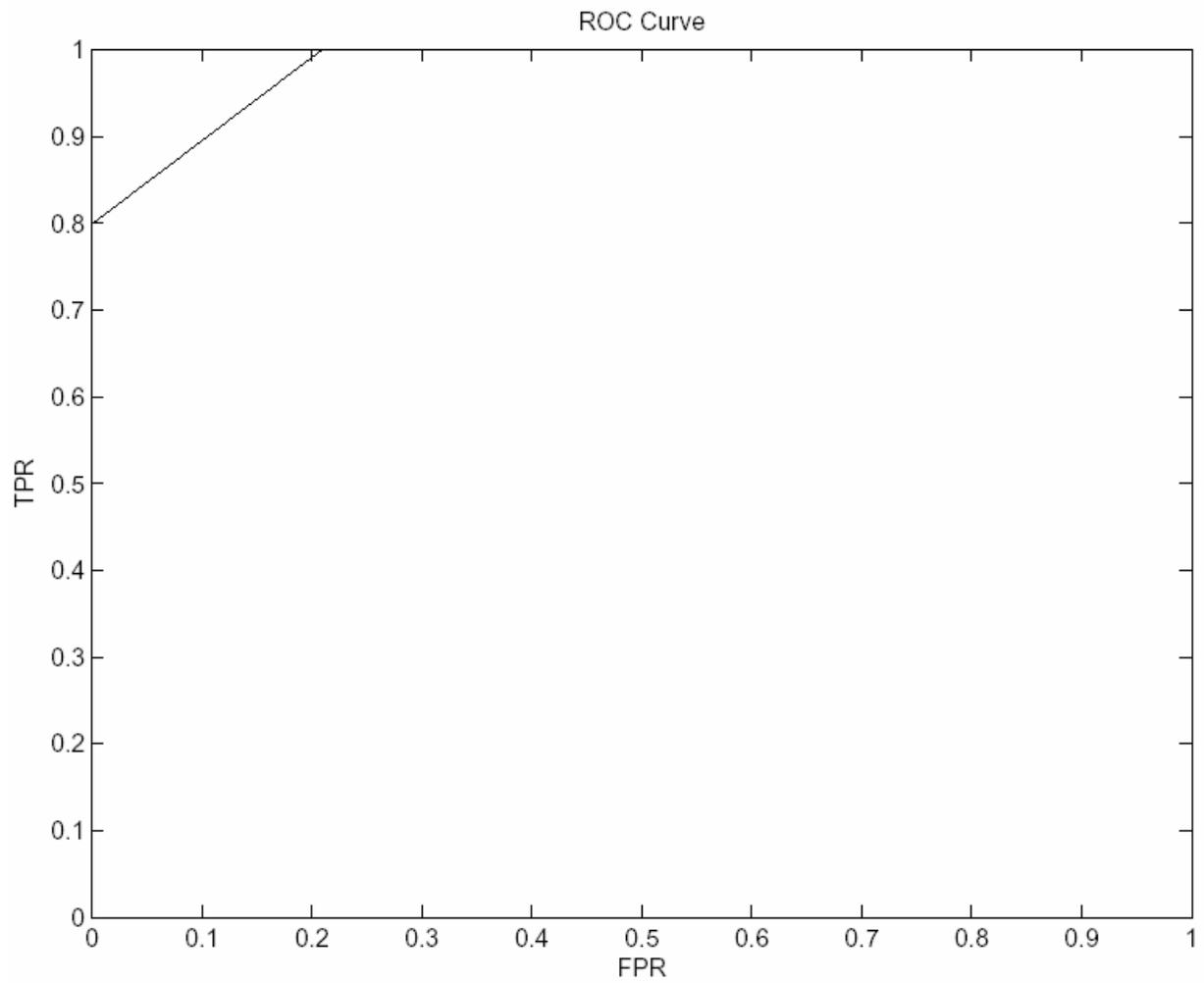

Fig. 3. The ROC curve of seizure detection. TPR = true positive rate and FPR = false positive rate. The area under the curve is $\approx 0.98$.

TABLE I

PATIENT DETAILS

| Patient | Sex | Age | Seizure type | H/NC | Origin | # seizures |
|---|---|---|---|---|---|---|
| 1 | F | 15 | SP,CP | NC | Frontal | 4 |
| 2 | M | 38 | SP,CP,GTC | H | Temporal | 3 |
| 3 | M | 14 | SP,CP | NC | Frontal | 5 |
| 4 | F | 26 | SP,CP,GTC | H | Temporal | 5 |

SP = simple parietal, CP = complex parietal, GTC = generalized tonic-clonic, H = hippocampal, NC = neocortical.

TABLE II

DETAIL OF THE PATIENT SPECIFIC PREPROCESSING

| Patient | Cut-off freq. | Montage |
|---------|---------------|---------|
| 1 | 100 Hz | Com. ref. |
| 2 | 50 Hz | Com. ref. |
| 3 | 100 Hz | Com. ref. |
| 4 | 50 Hz | Bipolar |